\documentclass[12pt]{iopart}

\usepackage{graphicx,color}% Include figure files
\begin{document}

\title[Unconventional quantum criticality]{Unconventional quantum criticality 
emerging as a new common language of transition-metal compounds, 
heavy-fermion systems, and organic conductors}

\author{Masatoshi Imada, Takahiro Misawa and Youhei Yamaji}

\address{Department of Applied Physics, University of Tokyo and JST-CREST, Hongo, Bunkyo-ku, Tokyo, Japan}
\ead{imada@ap.t.u-tokyo,ac,jp}
\begin{abstract}
We analyze and overview some of different types of unconventional quantum criticalities by focusing on two origins. One origin of the unconventionality is the proximity to first-order transitions. The border between the first-order and continuous transitions is described by a quantum
tricritical point (QTCP) for symmetry-breaking transitions. One of the characteristic features of 
the quantum tricriticality is the concomitant divergence of order-parameter and uniform fluctuations in contrast to the conventional quantum critical point (QCP). The interplay of these two 
fluctuations generates unconventionality.  Several puzzling non-Fermi-liquid properties in experiments 
are referred to be accounted for by the resultant universality
as in the cases of YbRh$_2$Si$_2$, CeRu$_2$Si$_2$ and $\beta$-YbAlB$_4$.
Another more dramatic unconventionality appears again at the border of the first-order and continuous transitions but in this case for  topological transitions such as metal-insulator and Lifshitz transitions.  This border, the marginal quantum critical point (MQCP), belongs to an unprecedented universality class with diverging uniform fluctuations at zero temperature. The Ising universality at the critical end point of the first-order transition at nonzero temperatures transforms to the marginal quantum criticality when the critical temperature is suppressed to zero.  The MQCP has a unique feature by a combined character of symmetry-breaking and topological transitions. In the metal-insulator transitions, the theoretical results are supported by experimental indications for V$_{2-x}$Cr$_x$O$_3$ and an organic conductor $\kappa$-(ET)$_{2}$Cu[N(CN)$_{2}$]Cl. Identifying topological transitions also reveals how non-Fermi liquid appears as a phase in metals.
The theory also accounts for the criticality of a metamagnetic transition in ZrZn$_2$, by interpreting it as an interplay of Lifshitz transition and correlation effects.  We discuss common underlying physics in these examples.  

\end{abstract}

\maketitle

\section{Introduction}
Quantum phase transitions and unconventional quantum phases are subjects of recent 
intensive studies.
In particular, a number of strongly correlated electron systems provide us with unconventional
types of quantum critical behaviors frequently accompanied by wide area exhibiting
non-Fermi-liquid properties in metals.
These are ranging from rare-earth compounds~\cite{Stewart,Lohneysen,Steglich}, transition-metal compounds~\cite{RMP} and 
organic conductors~\cite{KanodaJPSJ} implying the existence of universal underlying physics.

\begin{figure}[h!]
\begin{center}
\begin{tabular}{c}
\includegraphics[width=5cm]{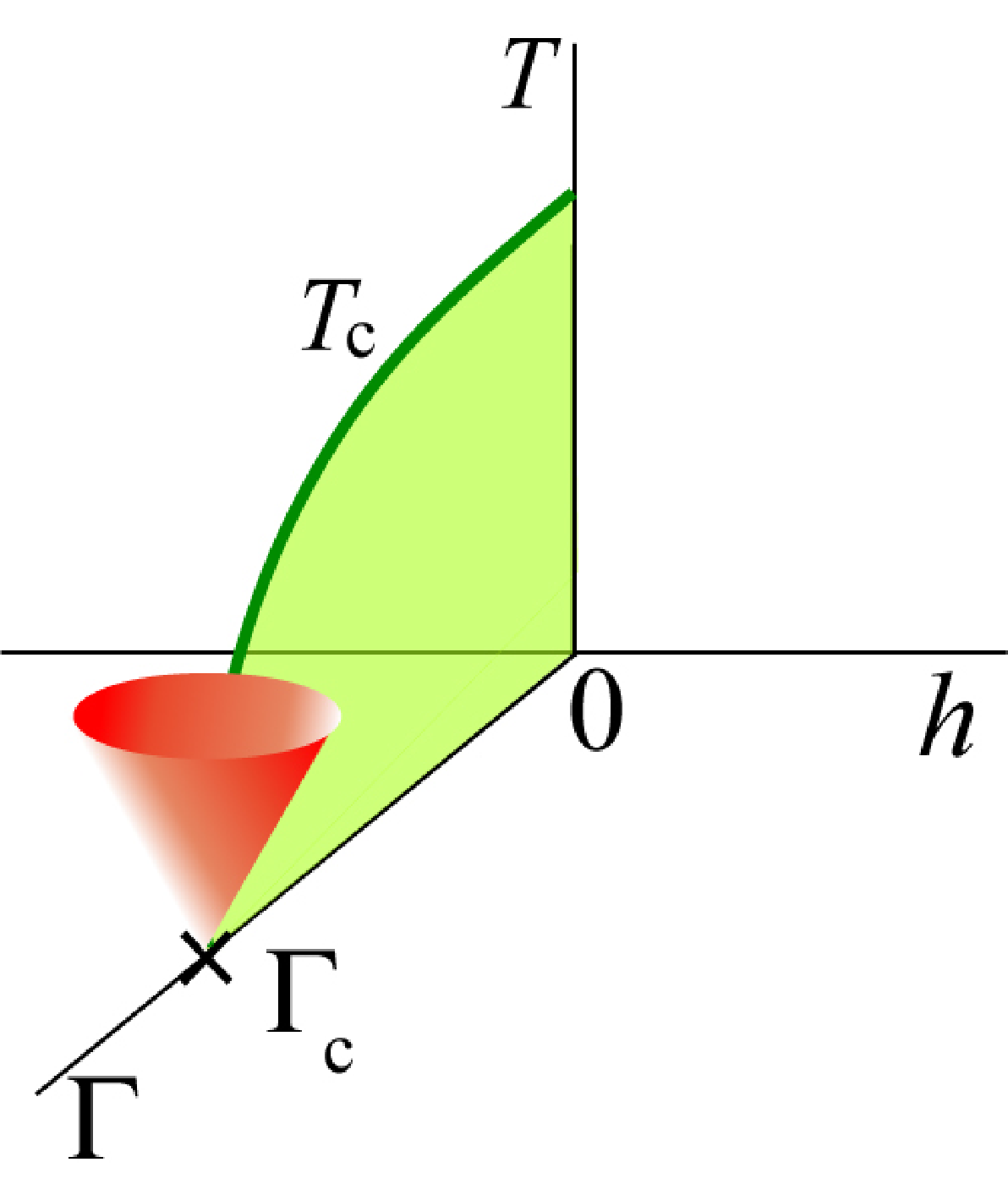}   
\includegraphics[width=7.8cm]{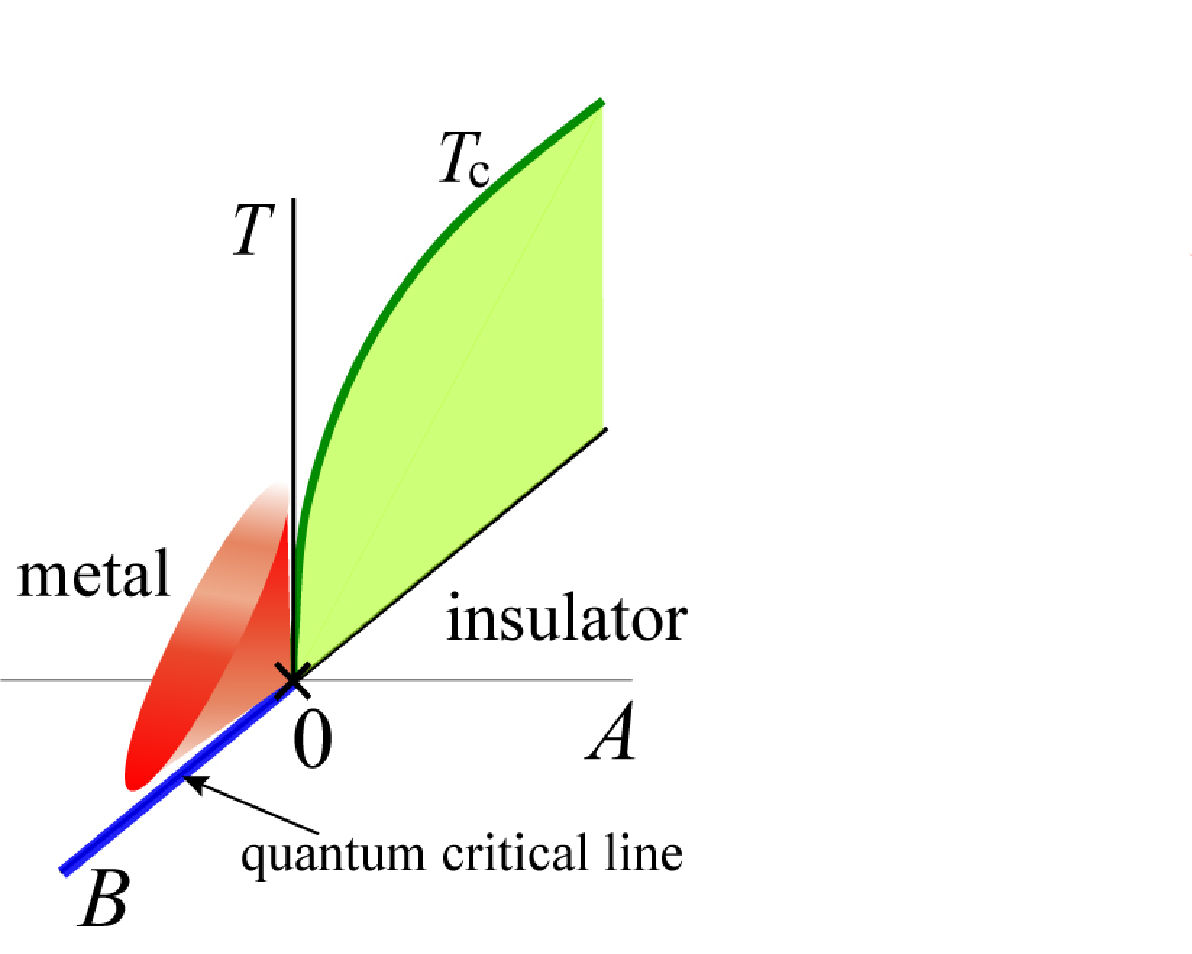}   
\end{tabular}
\caption{Phase diagram around conventional QCP (a)(left panel), and MQCP (b) (right panel for metal-insulator transition) in the parameter space of temperature $T$, fields to control transitions $h$ or $A$ and parameters to control quantum fluctuations $\Gamma$ or $B$. The cone structures schematically illustrate the quantum critical regions of the QCP (a) and MQCP (b) depicted by the crosses. First-order transitions occur when one crosses shaded (green) walls. Quantum critical line (bold (blue) line) in (b) represents continuous topological transition at $T=0$.}
\label{Fig1}
\end{center}
\end{figure}%
A prototype of quantum critical phenomena is found in the case where critical temperatures of 
spontaneous symmetry breaking such as magnetic ordering are suppressed to zero as we see in Fig.\ref{Fig1}(a) by enhancing some 
quantum fluctuations $\Gamma$. The parameter $\Gamma$ to enhance the quantum fluctuations of magnetic, charge or orbital orders 
in electronic systems is typically pressure or 
chemical doping, where itinerancy enhanced by these control parameters increases quantum fluctuations for the 
real space order realized by translational symmetry breaking.  Enhancing geometrical frustration effects
also increases quantum fluctuations.   
When the critical temperatures are suppressed to be low, low-energy and long-wavelength critical fluctuations of the order parameter start showing quantum mechanical character. 
In itinerant electron systems, this quantum critical fluctuation couples to low-energy quasiparticle excitations around the Fermi surface and leads to critical fluctuations qualitatively larger than the insulating case. This coupled case has been 
extensively studied by spin fluctuation theories developed by Moriya, Hertz and Millis\cite{Moriya,Hertz,Millis}.  
A feedback of the critical fluctuations to the quasiparticle excitations leads to non-Fermi-liquid behaviors typically observed in the region of a cone-shape structure as in Fig.\ref{Fig1}(a).  These standard spin fluctuation theories have been successful in explaining a number of experimental results on non-Fermi-liquid properties near the quantum phase transitions~\cite{Moriya,Takimoto}.
  
However, this standard theory has widely been challenged by recent progress in experiments~\cite{Stewart,Lohneysen,Steglich}. 
In some cases, critical exponents do not follow the prediction of the standard theory. 
In other cases, the critical region is unexpectedly large.
An important aspect ignored in the standard theory of quantum criticality is the interplay of itinerancy with localization effects caused by electron correlations.
Low-energy incoherent excitations on the verge of localization introduce a qualitatively new feature. 

In addition, novel quantum criticality in nature emerges when quantum phase transitions are not the consequences of the symmetry breaking. 
A completely different type of unconventional quantum phase transitions appears associated with 
{\it topological change} such as metal-insulator and Lifshitz transitions~\cite{Lifshitz} when they are combined with electron correlation effects as we describe in this paper. 
     
In this report, we first review understanding recently achieved for several different types of unconventional quantum criticalities.
Among various types of approaches for the unconventional quantum criticalities, we particularly focus on the cases where proximity to first-order transitions severely
modifies the conventional quantum criticality. 
This universal aspect offers a key for solving many puzzles and for understanding unconventional features in the experiments. 
The proximity to the first-order transitions is sometimes detected by signatures of 
spatial inhomogeneity and phase separations when the jump of the first-order transitions occurs in 
density under the fixed chemical potential.  This inhomogeneity is a subject of recent intensive studies in 
systems with competing orders, although we do not go into details of the issues of the inhomogeneity and phase separation.

\begin{figure}[h!]
\begin{center}
\begin{tabular}{c}
\includegraphics[width=10.0cm]{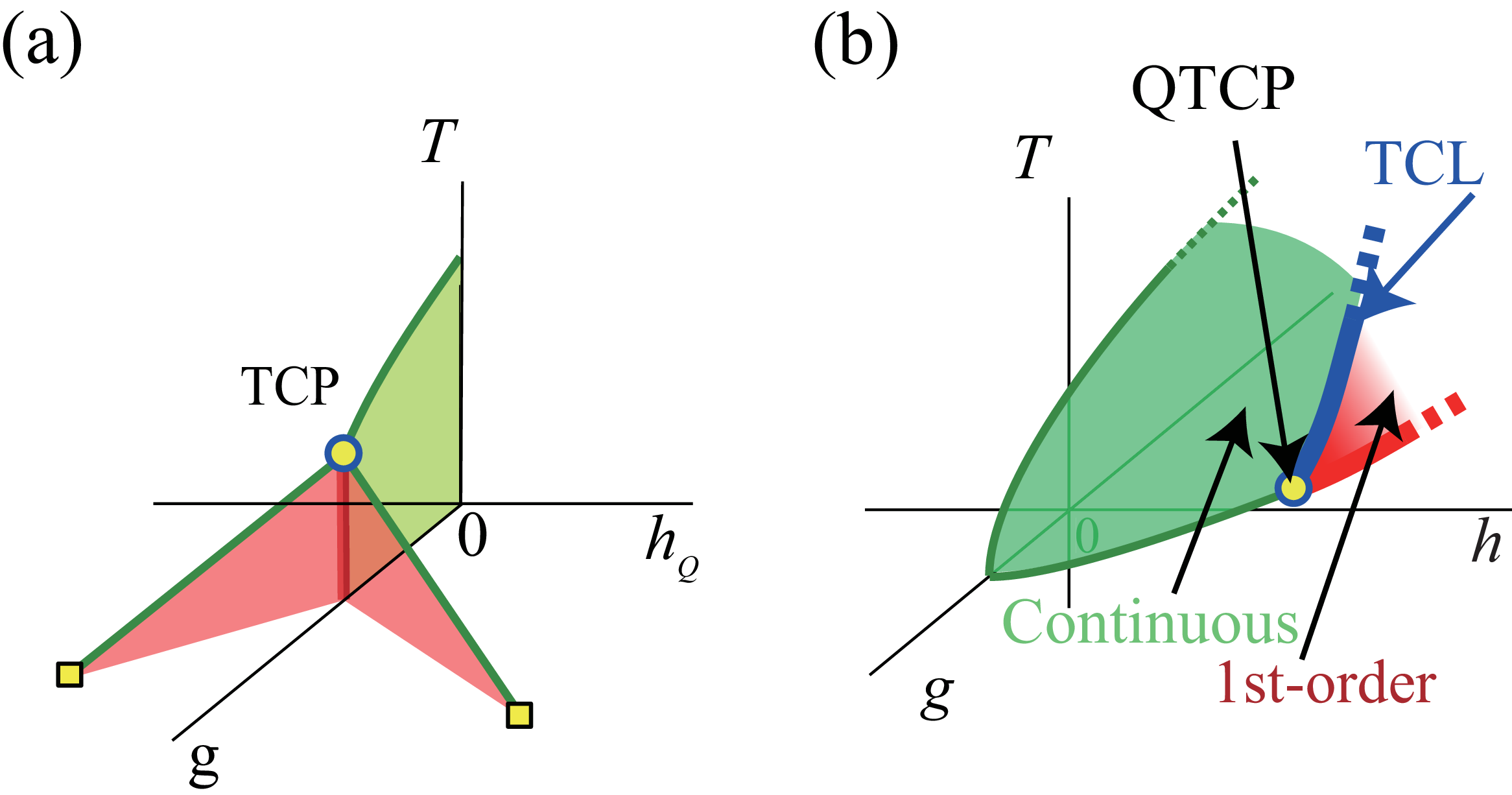}
\end{tabular}
\caption{
(a) Schematic phase diagram for continuous and first-order transitions together with classical TCP in parameter space of temperature $T$, quantum fluctuation $g$ and field $h_Q$ conjugate to order parameter $m_Q$.  Thin (green) lines represent critical lines of continuous transitions, while the route crossing the shaded sheets realize first-order transitions. At $h_Q=0$, the bold (red) line represent the first-order transition as well. The cirle is the classical TCP while the squares illustrate the QCP.  
(b) Global phase diagram with tricritical line (TCL) separating the surfaces of continuous 
[above TCL (green)] and the first-order [below TCL (red)] surfaces. 
Here $g$ represents a
parameter to control quantum fluctuations.
In YbRh$_{2}$Si$_{2}$, $g$ may correspond to pressure
measured from the ambient one and $h$ may be the uniform magnetic field.
The QTCP (circle) appears at $(g,H,T)=(g_{t},H_{t},0)$, namely the endpoint of TCL at $T=0$.
}
\label{fig2}
\end{center}
\end{figure}

A proximity to the first-order transitions in classical systems appears around the 
boundary between the continuous and first-order transitions called the 
tricritical point (TCP) \cite{tricritical} as is illustrated in Figs.~\ref{fig2}(a).
For example, at the
TCP of an antiferromagnetic transition under magnetic fields \cite{MisawaYamajiTricritical},
the jump of magnetization seen at the first-order transition is suppressed to zero, 
while a singular divergence of the magnetization slope as a function of magnetic fields appears.      
Then a unique feature of the TCP driven by magnetic fields is that the {\it uniform} magnetic susceptibility 
at zero wavenumber diverges in addition to the diverging order-parameter susceptibility at a nonzero wavenumber,
although it does not have tendency for the ferromagnetic order at all.
If the critical temperature of the TCP
is suppressed to zero, this transforms to a QTCP as we illustrate in a schematic phase diagram Fig~\ref{fig2}(b).
%Physics of the QTCP has recently been studied and its experimental relevance has been elucidated \cite{MisawaYamajiQTCP1}.
We discuss in Sec.\ref{Sect2} how an unconventional criticality appears in the case of the 
QTCP of the antiferromagnetic transition under magnetic fields \cite{MisawaYamajiQTCP1,MisawaYamajiQTCP2}, which has relevance in a number of $f$-electron systems including YbRh$_2$Si$_2$, CeRu$_2$Si$_2$ and $\beta$-YbAlB$_4$.  We also discuss possible origins of the proximity to first-order transitions, such as magnetic anisotropy and valence instability.   

The proximity to the first-order transition appears in a more dramatic way 
in case of the topological change.  
A simple example of the topological change is found in Fermi 
surface change such as Lifshitz transition and metal-insulator transition between a band insulator and a metal.
Although these topological transitions offer continuous quantum phase transitions, 
they do not cause any spontaneous symmetry breaking by themselves, while the critical phenomena are 
rather trivial in noninteracting systems. However, electron correlation introduces unprecedented effects.   
When the correlation effects become large, these transitions may become first-order transitions.    
The first-order transition continues to finite temperatures and is terminated at the finite-temperature critical point. 
Then this is well characterized by the conventional universality class of symmetry breaking,
where a similarity to the gas-liquid transition may be identified.  
The boundary between the first-order and conventional continuous topological transitions
illustrated in Fig.\ref{Fig1}(b) contains both characters of the topological and symmetry-breaking transitions~\cite{ImadaJPSJ,ImadaPRB,MisawaJPSJ,MisawaPRB}.
This point called the MQCP induces novel quantum critical phenomena around it. 
We clarify this novel physics in cases of metal-insulator transitions in Sec. 3.1 and
Lifshitz transitions in Sec. 3.2. 

The first-order transition and its proximity around the MQCP
are caused as the consequence of strong correlation effects, though the topological nature survives.
This compatibility is more deeply understood by the differentiation of quasiparticles in the momentum space.
The electron differentiation appears
in such a way that some particular part in the Brillouin zone shows strong correlation
effects with precursory insulating behavior while it leaves other part coherent as a small pocket of the Fermi surface.
The transition takes place as the topology change through the
vanishing pockets~\cite{Tahara,Sakai}. The topological nature
is better understood from the role of the zeros of
Green function (the poles of the self-energy), where the emergent zeros
nonuniformly destroys the large Fermi surface and
leaves the small pocket.

Through analyses on different types of the proximity to the first-order transition, in this paper,
we discuss underlying common physics with their relevances to experimental results for the unconventional quantum critical phenomena.
\section{Quantum tricriticality}
\label{Sect2}
In the classical Ginzburg-Landau-Wilson (GLW) scheme, the TCP is expressed by the $\phi^{6}$ theory~\cite{tricritical}.
The free energy $F$ is expanded up to the sixth order with respect to the scalar
order parameter $m_Q$ representing a spatial symmetry breaking at the wavenumber $Q$ as 
\begin{equation}
	F=rm_Q^{2}+um_Q^{4}+vm_Q^{6}-h_Q m_Q.
\label{phi4}
\end{equation}
If $u>0$, $r=0$ together with vanishing fields conjugate to the order parameter, $h_Q=0$ represent a conventional
Ising-type critical point.
If $u<0$, $u^2-3rv>0$ and $h_Q=0$, three-minima at $m_Q=0$ and $m_Q=\pm m_{Q0}$ with $m_{Q0} \equiv \sqrt{(-u + \sqrt{u^2-3rv})/3v}$ can represent the first-order transition
between $m_Q=0$ and $m_Q=\pm m_{Q0}$.
At $h_Q=0$, the first-order transition for $u<0$ and the continuous transition at $r=0$ for $u>0$ merge at $r=u=0$,
which determines the TCP.
Physics of the TCP has extensively been discussed for the mixture of
$^3$He and $^4$He as well as for antiferromagnetic transitions under magnetic fields~\cite{tricritical}.
A characteristic feature of the TCP is that not only the order-parameter susceptibility
diverges as $\chi_Q = (\partial^2 F /\partial m_Q^2)^{-1}=1/r \propto 1/(h-h_c)$ in this mean-field theory,
but also the uniform susceptibility $\chi_0 = (\partial^2 F /\partial h^2)$ diverges,
when the transition is controlled by uniform fields $h$ around the critical point $h=h_c$. In fact, $m_Q$ is scaled by $m_Q \propto r^{1/4}\propto |h-h_c|^{1/4}$ at $u=0$
and the resultant scaling of the free energy minimum $F\propto |h-h_c|^{3/2}$ leads to $\chi_0 \propto |h-h_c|^{-1/2}$. 
This indicates the scaling of the uniform magnetization $m_0-m_{0c} \propto \sqrt m_Q$ when $m_0$ is measured from the critical value $m_{0c}$.
We note that this diverging $\chi_0$ as $h\rightarrow h_c$ has nothing to do with the ferromagnetic 
tendency but is just the consequence of the tricriticality of the antiferromagnetic transition.
Because of the vanishing $u$ and $r$, the free energy becomes flattened around $m_Q=0$ and hence fluctuations around the critical point become large in general.   It also causes the diverging uniform susceptibility.

When the tricritical temperature is suppressed by quantum fluctuations, the QTCP
appears.  In this case, when the transition occurs in metallic phases, the critical fluctuations of bosons associated with the 
order-parameter fluctuations couple to 
the low-energy quasiparticle excitations near the Fermi surface similarly to the conventional QCP
in metals~\cite{Moriya,Millis}.  However, in the case of the QTCP, it has features qualitatively different from the conventional quantum criticality already known in the classical case.  To elucidate this, we have proposed a spin fluctuation theory
for the antiferromagnetic QTCP~\cite{MisawaYamajiQTCP1,MisawaYamajiQTCP2}.
As in the classical case, under magnetic fields, ``ferromagnetic" quantum critical 
fluctuations develop around the antiferromagnetic QTCP in addition to
antiferromagnetic fluctuations, which is in sharp contrast
with the conventional antiferromagnetic QCP.
For itinerant electron systems,
it has been shown that the temperature dependence of
critical magnetic fluctuations around the QTCP
is given as $\chi_{Q}\propto T^{-3/2}$ ($\chi_{0}\propto \sqrt{\chi_Q} \propto T^{-3/4}$) 
at the antiferromagnetic (ferromagnetic) wave number $q=Q$ ($q=0$).
The convex temperature dependence of $\chi_{0}^{-1}\propto T^{3/4}$ is 
the characteristic feature of the QTCP, which should not be
seen in the conventional spin fluctuation theory for the ferromagnetic transition
because the exponent $3/4$ is smaller than unity for the Curie law.
The same scaling leads to the singular magnetization process $m \propto |h-h_c|^{1/2}$.
It should be noted that these critical exponents are completely different 
from the conventional quantum  criticality.
\begin{figure}[h!]
\begin{center}
\includegraphics[width=10cm]{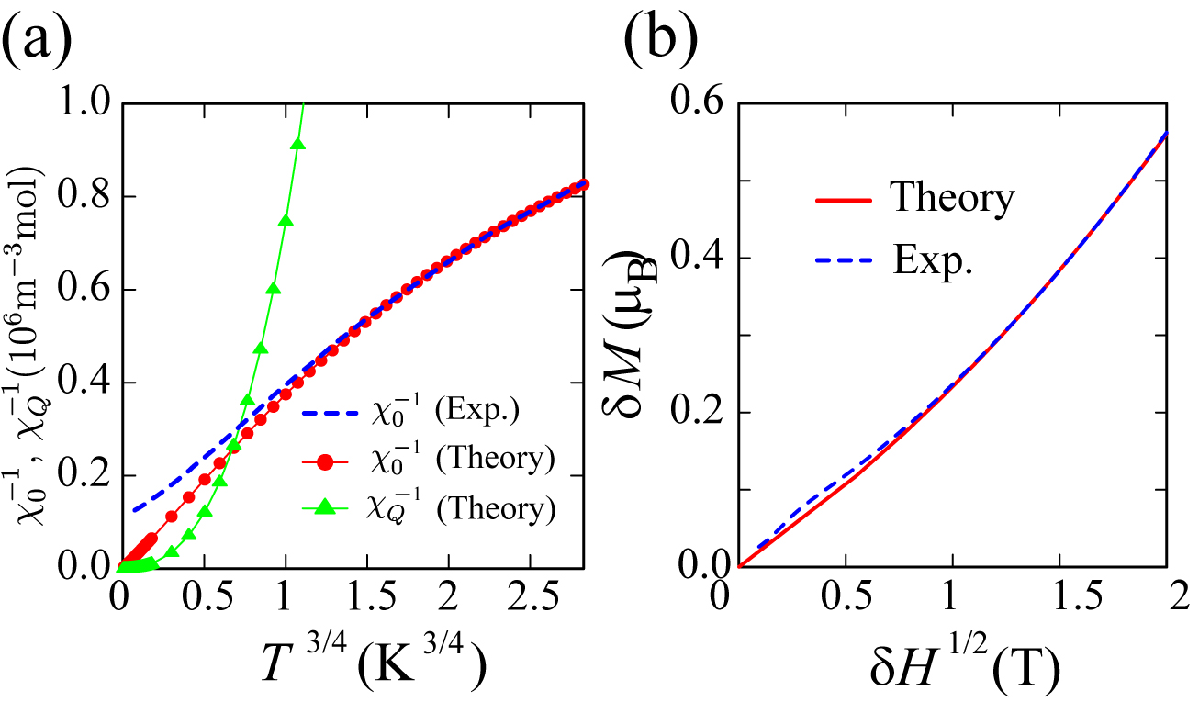}   
\caption{ (a)  Temperature dependence of inverse uniform magnetic susceptibility $\chi_{0}^{-1}$ 
for YbRh$_{2}$(Si$_{0.95}$Ge$_{0.05}$)$_{2}$
at $H=0.03$ T reported in Ref.~\cite{Gegenwart_1} illustrated as broken (blue) curve compared with
numerical result of spin fluctuation theory for the QTCP~\cite{MisawaYamajiQTCP1,MisawaYamajiQTCP2}
shown as solid (red) curve with filled circles.
The deviation at low temperatures appears because the experimental parameters are deviated from the QTCP. 
Solid (green) curve with triangles represents the theoretical
$\chi_{Q}^{-1}$.
(b) Magnetic field dependence of magnetization (broken (blue) curve) 
for YbRh$_{2}$(Si$_{0.95}$Ge$_{0.05}$)$_{2}$
at $T=0.09$ K reported in Ref.~\cite{Gegenwart_1}
%~\cite{Gegenwart_1} 
compared with the QTCP theory (solid (red) curve)~\cite{MisawaYamajiQTCP1,MisawaYamajiQTCP2}. 
$\delta M$ ($\delta H$) represents the magnetization (magnetic field)
measured from the critical value.
We estimate the experimental critical magnetic field $H_{c}$ (magnetization $M_{c}$)
as 0.027 T (0.004 $\mu_{\rm B}$).}
\end{center}
\label{Fig2}
\end{figure}%

It has been shown that physics of the QTCP accounts for several unconventional features of the quantum criticalities and 
non-Fermi-liquid properties observed experimentally in heavy-fermion system{s}
such as YbRh$_{2}$Si$_{2}$, CeRu$_{2}$Si$_{2}$, and $\beta$-YbAlB$_{4}$~\cite{MisawaYamajiQTCP1,MisawaYamajiQTCP2}.
For YbRh$_{2}$Si$_{2}$, 
the QTCP successfully reproduces
quantitative behaviors of
the experimental ferromagnetic susceptibility $\chi_0 \propto T^{-0.6}$ 
by an appropriate choice of the phenomenological parameters.
In fact, a crossover from $\chi_{0}\propto T^{-3/4}$
to $\chi_{0}\sim T^{-0.6}$ with elevated temperatures predicted by the theory of the QTCP
quantitatively reproduces the 
experimental result as seen in Fig.\ref{Fig2}(a).
The deviation at low temperatures is ascribed to the deviation of the experimental parameters 
from the right QTCP.
Figure \ref{Fig2}(b) also shows that the magnetization curve follows the prediction of the 
quantum tricriticality $m \propto |h-h_c|^{0.5}$. 

The quantum tricriticality also reproduces singularities of  
other physical properties such as specific heat, nuclear magnetic relaxation time $1/T_{1}T$,
and the Hall coefficient observed for YbRh$_{2}$Si$_{2}$.
A simple argument\cite{Coleman} predicts that  
the Hall coefficient $R_{\rm H}$ is scaled by 
$R_{\rm H}\propto {m_Q^{\dagger}}^{2}$, while as is mentioned above,@
$m_Q\propto |h-h_{c}|^{1/4}$ holds.
Therefore, Hall coefficient is scaled by 
$|h-h_{c}|^{1/2}$ near the QTCP. 
This scaling indicates that the Hall coefficient 
shows a singular change near the QTCP. 
If the QCP in YbRh$_2$Si$_2$ is located on the side of weak first-order phase transitions, 
the Hall coefficient changes even jumps at $T=0$. 
A steep increase of $R_H$ in the experiment~\cite{Paschen} at low temperatures 
is consistent with the present prediction. 

Under magnetic fields $h>h_{c}$,
two characteristic temperature scales are suggested in YbRh$_{2}$Si$_{2}$~\cite{Gegenwart_1,Gegenwart_sc};
Below one scale ($T_{\rm LFL}$), the Landau
Fermi liquid becomes satisfactory while around the other scale $T^{*}$,
$\partial R_{\rm H}/\partial H$, $\partial \rho/\partial H$, and $\chi_{0}$
have peaks.
We note that the coexistence of the antiferromagnetic and ferromagnetic
fluctuations is a possible origin of the observed two energy scales:
$T^{*}$ is interpreted as the energy scale
where the uniform fluctuation $\chi_{0}$ starts saturating where the response
to the uniform magnetic field shows an anomaly.
The other is $T_{\rm LFL}$ below which the antiferromagnetic fluctuations 
saturate.  Since the growth of the antiferromagnetic and uniform flutuations 
both destroys the Fermi liquid scaling, the real Fermi liquid shows up 
only when both of them saturates, namely only below the lower scale $T_{\rm LFL}$.  
Since both FM and AFM fluctuations diverge at the QTCP, 
two energy scales $T^{*}$ and $T_{\rm LFL}$ vanish at the QTCP.
These are consistent with the experimental result. 

Recently, it has been pointed out that the specific heat exponent for the tricriticality
scales as $C \propto |T-Tc|^{-1/2}$~\cite{Popov,MisawaYamajiQTCP2} This is consistent with the experimental observation~\cite{Steglich2}. 

The proposal for the proximity of the first-order transition and the tricriticality is also supported from the real existence of the first-order transition under pressure~\cite{Knebel}, where the jump of the resistivity is clearly seen at 2.3 GPa under the magnetic field $H \parallel c$ around 2 T. Since the tricritical temperature is around 0.5K at this pressure, while the transition is always continuous at ambient pressure, the QTCP has to show up between these two pressures.  The physics of QTCP can be more definitely tested at this anticipated QTCP and we propose experiments under the tuning of pressure.  
Recently, effects of chemical pressure have been examined by substituting Co for Rh~\cite{Klingner}.  It suggests a complex phase diagram: For a small concentration of Co up to $x=0.28$ for Yb(Rh$_{1-x}$Co$_x$)$_2$Si$_2$, the transition becomes broadened without an indication of the first-order transition under magnetic fields perpendicular to the $c$ axis. On the other hand, a first-order transition is signaled for large $x\sim 0.68$ under magnetic fields parallel to the $c$ axis and even in the absence of magnetic fields. The absence of the first-order transition at small $x$ was claimed~\cite{Klingner} to contradict the experiment under the hydrostatic pressure~\cite{Knebel}, if the chemical pressure and the hydrostatic pressure could be mapped.  However, since the first-order transition under the hydrostatic pressure is observed only in magnetic fields parallel to the $c$ axis, it is desired to examine the chemical pressure effect under the same condition, because the mechanism may involve the effect of magnetic anisotropy as we will discuss below.   

For CeRu$_{2}$Si$_{2}$~\cite{SuzukiH} and $\beta$-YbAlB$_{4}$~\cite{Nakatsuji} as well,
the quantum tricriticality is a presumable 
origin of the anomalous diverging enhancement of 
the uniform susceptibility observed in these materials~\cite{MisawaYamajiQTCP2}.

Let us discuss mechanisms of generating first-order transitions.  It is known that YbRh$_{2}$Si$_{2}$ and $\beta$-YbAlB$_{4}$ have a strong magnetic anisotropy. In fact, the QCP for YbRh$_{2}$Si$_{2}$ is realized at $\sim 0.06$T for the magnetic field perpendicular to the $c$ axis while 0.6T is required for the field parallel to the $c$ axis.  In classical metamagnetic systems, the single-site anisotropy commonly causes a first-order (metamagnetic) transition from an antiferromagnetic phase to a spin-flipped paramagnetic phase under magnetic fields as in the case of FeCl$_2$~\cite{FeCl2}. The first-order transition under pressure is so far observed only in magnetic fields parallel to the $c$ axis~\cite{Knebel}, which may be related to this type of the mechanism.

Another possible origin driving the first-order transition is the valence instability in the $f$-electron systems. The $f$ electrons located near the Fermi level can hybridize with the conduction electrons $c$ leading to the emergence of heavy-mass quasiparticles through the Kondo effect. The valence of the $f$ electrons may abruptly change through the shift of the $f$ electron level relative to the conduction electrons $c$ and/or the competition among the conduction bandwidth, the $c$-$f$ hybridization, and the atomic $f$-$f$ as well as $c$-$f$ electron correlations. This transition may lead to the formation/destruction of the $f$ electron local moment. This dominates physics of the $\gamma$-$\alpha$ transition of Ce~\cite{Koskenmaki}.  The valence transition sometimes occurs as a first-order transition.  If it happens as the first-order jump, the universality of the valence transition is described by the type of the gas-liquid transition with a finite temperature critical point characterized by the Ising universality.  When this critical temperature is suppressed to zero, a conventional QCP that is equivalent to that in Fig.\ref{Fig1}(a) appears.

Now, if the valence transition equivalently described by Fig.\ref{Fig1}(a) occurs simultaneously with the magnetic transition, this valence critical point ($T_c$ line in Fig.~\ref{Fig1}(a)) may be transformed to the tricritical point (tricritical line).  This is because even outside the shaded (green) sheet in Fig.\ref{Fig1}, the left and right sides of the shaded sheet have to be distinguished by the symmetry difference of the simultaneous magnetic transition. This means that the shaded (green) sheet continues to form a sheet of continuous transition beyond the $T_c$ line. This is nothing but the appearance of the tricritical line that replaces the $T_c$ line in Fig.~\ref{Fig1}. In this sense, the quantum tricriticality may capture relevant physics even when valence transition is on the verge of the magnetic first-order transition.

A closely related idea is the localization transition of the $f$ electron through the Kondo collapse (Kondo breakdown), namely the switch-off of the $c$-$f$ hybridization disconnecting the $f$ electrons from the conduction band, for example, through decreasing pressure. The Kondo collapse may take place either as a continuous or a first-order transition.  When it is combined with a magnetic transition, the quantum tricriticality similar to the case of the valence transition may occur.  Therefore, the quantum tricriticality may capture relevant physics even when the valence transition or the so-called local quantum criticality~\cite{Coleman} is on the verge of the magnetic first-order transition.

On the other hand, if the valence transition or Kondo breakdown involves a topological change of the Fermi surface, the transition may have a structure essentially described by Fig.\ref{Fig1}(b). In fact, this type of universality will be described in the next section. Here we note that the Kondo breakdown interpreted by the orbital-selective Mott transition of the $f$ electrons indeed suggests the applicability of the universality discussed in the next section~\cite{Pepin,Leo}. In this sense, the present classification and concept of unconventional quantum criticality offer a useful and general scheme for describing $f$-electron systems.

\section{Marginal quantum criticality}
\label{Sect3}
The proximity to the first-order transition appears in a different way when the underlying quantum criticality 
belongs to a different class.
An intriguing issue is the quantum phase transition that is driven not by spontaneous symmetry breaking 
but by some topological change. 
In general, quantum phase transitions caused by change in topological number occur 
in a wider class of phenomena including
the quantum Hall effects~\cite{quantumHall} and topological insulators~\cite{KaneMele}.
A simple example is seen in the change in topology of the Fermi surface.
In this section, we visit two examples of this category.
\subsection{Metal-insulator transition}
The first example is metal-insulator transitions driven by electron correlation effects.
Such a well known example is the Mott transition~\cite{RMP}.

In general, the metal-insulator transitions in weakly correlated systems take place either as
the transition between Fermi liquids and band insulators or as the Anderson transitions driven by disorder.
Both of these cases are essentially identified as the transitions at zero temperature.
In these cases, unless some other origins such as a 
structural phase transition drive the metal-insulator transition and forces discontinuous 
changes in the band structure through a strong electron-lattice coupling, the phase transitions are basically of continuous type. 

However, when electron correlation effects play a role, transitions may frequently appear as first-order transitions.
Even without a relevant coupling to the lattice, it is now believed that first-order transitions generically 
appear in nature.  When the bandwidth is controlled either by pressure or chemical pressure realized through chemical 
substitutions, such first-order metal-insulator transitions are ubiquitously observed.
A well known example is found for V$_{2-x}$M$_x$O$_3$ with M=Ti or Cr. In these compounds, 
the first-order transition terminates at around 350K identified as the critical point~\cite{McWhan1971}.  
The universality class of this critical point has been studied carefully from the 
conductivity exponent and it has been established that it belongs to the Ising universality class~\cite{Limelette}.
Although the order parameter is not trivial, the transition is essentially described by the symmetry-breaking type.
It indicates that this Mott transition is equivalent to the gas-liquid transition that is known to be described by the Ising universality. 
In fact, the natural order parameter is the carrier density as in the case of the density in the gas-liquid transition.
The Ising universality also implies that the transition is described by the conventional GLW scheme~\cite{GinzburgLandau,Wilson}. 
This has immediately raised a fundamental question about the nature of this class of metal-insulator transition,
because neither the metal to band-insulator transition nor the Anderson transition are known to belong 
to this universality class.  

A successful phenomenological description is constructed starting from the free energy for the band-insulator
metal transition.  When the Fermi level crosses the bottom (top) of the band dispersion $\varepsilon (k) \propto k^z$ for electrons (holes), the free energy (or the energy at $T=0$) of noninteracting electrons in the metallic phase is given by 
   \begin{equation}
    F_0 \propto \int_0^{E_F} d\varepsilon \varepsilon D(\varepsilon) \propto X^{(d+z)/d},
    \label{hyperscaling2}
  \end{equation}
where $E_F$ is the Fermi energy and $D(\varepsilon)\propto \varepsilon^{(d-z)/z}$ is the density of states
for spatial dimension $d$. The free carrier density is taken as the natural order parameter and defined as $X \ge 0$. 
In case of the generic dispersion expanded as $\varepsilon (k) \propto ak^2+bk^4+\cdots$, the free energy in the grand canonical ensemble is reduced to 
   \begin{equation}
    F_0 \propto -\mu X + c_2 X^{(d+2)/d}+ c_4 X^{(d+4)/d}+ \cdots ,
    \label{hyperscaling3}
  \end{equation}
with constants $c_2$ and $c_4$ and the chemical potential $\mu$ for the carrier. 
In the insulating side, $F_0=0$ is trivially satisfied.
Now the interaction energy may be introduced by the form of effective two-body interaction of carriers scaled as
   \begin{equation}
    F_1 \propto X^{2}.
    \label{hyperscaling4}
  \end{equation}
Then the total energy (free energy) in the metallic phase is given by 
   \begin{eqnarray}
    F &=&F_0+F_1 = -\mu X + vX^2 + c_2X^{(d+2)/d} + c_4X^{(d+4)/d} +\cdots .
    \label{hyperscaling5}
  \end{eqnarray}
For $d=1$, it turns out that the second lowest order term is proportional to $c_2$ 
while it is $v$ for $d=3$. Two dimensional systems have a unique feature because 
the terms proportional to $c_2$ and $v$ are the same order. As we see below this leads to an unconventional
dynamical exponent $z=4$ for the critical point $\mu=v+c_2=0$.  
It is now clear that though it has an expansion in terms of $X$, 
this form of the free energy does not follow the simple GLW scheme in any dimension. 
In fact when one moves the chemical potential as the control parameter, the expansion (\ref{hyperscaling5}) is justified only in the metallic phase for larger $\mu$, while the free energy in the insulator side described by smaller $\mu$ has the minimum zero always at $X=0$, that means this piece-wise analytic character does not allow the analytic expansion of the free energy itself in contrast to the case of Eq.(\ref{phi4}).  
This breakdown originates from the fact that the metal-insulator transition between $X=0$ and $X>0$ is dominated by the topological character of the Fermi-surface pocket
on the verge of the transition.  The transition is not originally described by any type of the symmetry breaking
but by the topological change in the ground state, where the singular form of the density of states $D$ determines
the nonanalytic expansion.

Let us focus on the two dimensional case, where Eq.(\ref{hyperscaling5}) is reduced to 
   \begin{eqnarray}
    F &=& A X + BX^2 + CX^{3} +\cdots .
    \label{hyperscaling5-2}
  \end{eqnarray}
 Then the effective interaction (quadratic term) is 
proportional to $B$. When $B$ is positive, the metal-insulator transition occurs as a continuous transition
by controlling $A$ through zero.  
However, if the effective interaction $B$ is driven to a negative value, a first-order transition occurs at a certain $A>0$. 
The first-order transition is transformed to the continuous one at the MQCP determined by $B=0$ and $A=0$.  The universality class of the MQCP is unconventional and is characterized by the critical exponents in the standard notation as 
   \begin{equation}
  z=4, \alpha=-1, \beta=1, \gamma=1, \delta=2, \nu=1/2$ {\rm } and $\eta=0.
    \label{exponents}
  \end{equation}

For the parameter for the first-order transition, $B<0$, the jump of $X$ obviously continues to nonzero
temperatures and the jump terminates at the critical point. For the critical point at $T>0$, the free-energy form (\ref{hyperscaling5-2})
is no longer valid, because the singular form of $D$ is immediately smeared out by the Fermi distribution 
at $T>0$.  Then the double minima form of the free energy expansion is regular as 
   \begin{equation}
    F = -\mu X + A'X^2 + B'X^4 +\cdots .
    \label{hyperscaling6}
  \end{equation}
which leads to the conventional Ising universality class~\cite{ImadaJPSJ,ImadaPRB,MisawaJPSJ,MisawaPRB}.
Now it has turned out that the MQCP is sandwiched by the topological quantum critical line for $B>0$ at $T=0$
and the Ising critical line at $T>0$ as is sketched in Fig.~\ref{Fig1}(b). 
The unconventionality arises from this emergent character, 
which appears at the marginal point between
the Ising-type symmetry breaking and the topological transition of the 
Fermi surface at zero temperature~\cite{ImadaJPSJ,ImadaPRB,MisawaJPSJ,MisawaPRB}. 

It has been shown that even Hartree-Fock approximations of 
an extended Hubbard model on square lattices
are capable of such metal-insulator transitions 
with  unusual criticality under a preexisting symmetry breaking~\cite{MisawaJPSJ,MisawaPRB}.
In this case, the above free-energy expansion can indeed be obtained from microscopic models analytically as well as numerically. 
The obtained universality perfectly agrees with the above critical exponents
and with a number of numerical results beyond the mean-field level as well~\cite{Furukawa,Assaad},
implying that 
the preexisting symmetry breaking assumed in the Hartree-Fock study is not necessary for this unconventional universality.
Furthermore, examinations of fluctuation effects indicate that the critical exponents 
remain essentially exact beyond the mean-field level except for the possible logarithmic correction, 
because the upper critical dimension 
$d_{c}$ is given by~\cite{MisawaPRB} 
	\begin{equation}
		d_{c}=\frac{\gamma+2\beta}{\nu}-z 
			=2.	
	\end{equation}
 
The critical exponents identified by the conductivity measurements
for V$_{1-x}$Cr$_x$O$_3$ at finite temperatures~\cite{Limelette} 
agree with the Ising exponents derived here. On the other hand, 
an organic conductor $\kappa$-(ET)$_{2}$Cu[N(CN)$_{2}$]Cl has a low-temperature critical point 
of the metal-Mott-insulator transition, where the same conductivity measurement has revealed unconventional exponents 
$\beta\sim 1,\gamma\sim 1$
and $\delta \sim 2$~\cite{Kanoda}. These observations perfectly agree with the universality class of the MQCP~\cite{Papanikolaou}. 
In fact, the exponents for the MQCP at $T=0$ survives as crossover exponents and dominates even at
nonzero-temperature critical point~\cite{MisawaPRB} as in the case of $\kappa$-(ET)$_{2}$Cu[N(CN)$_{2}$]Cl.
The careful experimental results 
in (V,Cr)$_2$O$_3$ and $\kappa$-ET-type organic conductor $\kappa$-(ET)$_{2}$Cu[N(CN)$_{2}$]Cl support the existence of 
the present marginal quantum criticality connecting the Ising-type critical line and the topological quantum critical line. 
The thermal expansion coefficient $\alpha(T) =l^{-1}dl/dT$ of $\kappa$-(ET)$_{2}$Cu[N(CN)$_{2}$]Cl has also been reported to have an anomaly~\cite{Souza}. Since the data seem to strongly depend on the form of the distribution of the transition temperature, it is difficult to determine the exponent of the singularity quantitatively.  Even though, this anomaly
can be qualitatively interpreted by the singularity of the mobile carrier density, because the lattice expansion linearly couples to the carrier density~\cite{Papanikolaou}. If the present quantum criticality also dominates for a uniform system, we expect the exponent of $\zeta=-1/2$ for $\alpha(T) \propto (T-T_c)^{\zeta}$, because 
the temperature axis crosses the transition from metallic to insulator side and $\zeta$ generically probes $1/\delta -1$. 

To obtain a direct evidence of the critical exponents, it is desirable to precisely determine the singularity of the carrier density. Finding a system with a lower critical temperature and revealing behaviors of Fermi surface topology are also very important challenge left for the future in this fundamental issue of the quantum Mott transition. 

It is highly suggestive in terms of possible superconducting mechanisms 
that the MQCP emerges on the verge of the effective interaction of carriers driven to be
attractive.  Although the Ising critical point appears when the effective interaction is driven to be attractive 
as in the case of the liquid-gas transition, for the superconducting pairing, the attractive interaction leading to
the Cooper pairing has to be realized in the region of the Fermi degeneracy. This is only possible around the MQCP~\cite{ImadaJPSJSuper}. 

The present results imply that the metal-insulator transition is governed by a topological change
in the Fermi surface with shrinkage (or emergence) at selected momentum points even when the interaction effects dominate.
This is different from other types of scenario such as that from the dynamical mean-field
 approximations, where the metal-insulator transition is governed instead by the vanishing 
 renormalization factor $Z$ and a large Fermi surface is retained even on the verge of the transition.
On the verge in the metallic side, the topological character suggests that the Fermi surface is 
reduced to small pockets which violates the Luttinger sum rule.
If the system undergoes a Lifshitz transition from a large to a small
Fermi surface, this violation is allowed in the side of small pockets.
 The dynamical mean-field theory improved by including the momentum dependence of the self-energy
indeed suggests the existence of 
 such a Lifshitz transition~\cite{Hanasaki}, and
the resultant shrinking small Fermi pockets in the absence of the translational symmetry breaking~\cite{Sakai,Zhang}.

In the mechanism of realizing the topological character of the metal-insulator transition, it has turned out that the zero of the single-particle Green function plays an important role~\cite{Sakai,Dzyaloshinskii,Rice,Stanescu}. The single-particle Green function is defined as 
	\begin{equation}
		G(k,\omega) = \frac{1}{\omega -\varepsilon (k) - \Sigma(k,\omega)},	
	\end{equation}  
where $\Sigma$ is the free-energy and $\varepsilon$ is the dispersion of the noninteracting part.
Now, when $\omega$ is largely negative, ${\rm Re}G<0$ must always be satisfied, while ${\rm Re}G>0$ for largely positive $\omega$.
Then a sign change has to occur at an intermediate value of $\omega$ at least once. In metals, the sign change indeed occurs through ${\rm Re}G=\pm \infty$ obtained from the pole of the Green function $\omega=\varepsilon (k) + \Sigma(k,\omega)$, which determines the Fermi surface of metals at $\omega=0$.  However, the sign change may also occur through ${\rm Re}G=0$, which corresponds to the pole of the self-energy $\Sigma$.  In fact, the above sign change in $G$ has to occur between $\omega \gg 0$ and $\omega \ll 0$ even in insulators while the Fermi surface does not exist in insulators at all. The sign change in insulators actually occurs through the zeros of the Green function. 
Since $\varepsilon(k) < 0$ at the Brillouin zone center ($\Gamma$ point) while it is positive at the zone boundary (for example, at $(\pi,\pi)$ in 2D systems), ${\rm Re} G$ has to change the sign between these two point at $\omega=0$, which determines the Fermi surface in metals. Even in the Mott insulator, this sign change equally has to occur, the only way of which is through the zeros of $G$. Therefore, a zero surface has to cross the Brollouin zone at $\omega=0$~\cite{Rosch}.
Since the self-energy is divergent at the zeros, the perturbation expansion obviously breaks down at the zeros.  For the continuous metal-insulator transitions, the poles cannot be replaced with the zeros abruptly at the transition, while poles completely disappear in the insulator side at $\omega =0$ and the zeros dominate.  This means that the emergence of the zeros has to occur already in the metallic side with a progressive replacement of the poles with zeros.  
When a topological transition ascribed to the interaction effects such as a transition of zeros emergence or a Lifshitz transition occurs at $\omega=0$, this is the point of the breakdown of the Fermi liquid in the strict sense, and non-Fermi liquids show up, because the system is not adiabatically connected to the noninteracting system any more.  It is clear that the breakdown of the Fermi liquid occurs in an inhomogeneous way in the Brillouin zone depending on the distance from the zeros.  Near the zeros, the quasiparticles become more incoherent because of the enhanced $\Sigma$ and it introduces the differentiation of electrons.  It has been shown that such a differentiation of electrons eventually leads to a breakup of the original large Fermi surface by the interference and penetration of the zeros to the poles. After the destruction of the original Fermi surface caused by the zeros, the remaining part of the Fermi surface becomes pockets and the pockets shrink to disappear at the {\it topological} metal-insulator transition. From this clarification, it turns out that the topological character of the metal-insulator transition clearly leads to the emergence of a non-Fermi liquid as an extended phase realized by the appearance of the zeros.  Effects of zeros and differentiation of electrons on the thermodynamic as well as transport properties are not fully understood yet and left as an intriguing issues.

%In $d=3$, the critical exponent of the MQCP is given from Eq.(\ref{hyperscaling5}) as $z=2, \alpha=-3, \beta=2, \gamma=1, \delta=3/2, \nu=1/2$ and $\eta=0$. This is the same universality class as one type of the Lishitz transitions as we describe below. 

\subsection{Lifshitz Transition}
Topological change in the Fermi surface called the Lifshitz transition has originally been studied for noninteracting electrons~\cite{Lifshitz}. Recently, interaction effects on the Lifshitz transitions have been systematically studied~\cite{YamajiLifshitz}. When the electron Coulomb interaction is switched on, the first-order transition may appear similarly to the case of the metal-insulator transition discussed in the previous section.  The marginal point between the continuous and first-order transition lines again appears as the MQCP.   When the Lifshitz transition takes place in the ferromagnetic phase and the first-order transition is driven by magnetic fields, it appears also as the metamagnetic transition. 

Similarly to the case of metal-insulator transitions described by Eq.(\ref{hyperscaling5}), when a Fermi pocket vanishes at a Lifshitz transition, the free energy can be expanded by the magnetization $\Delta m$ and magnetic field $\Delta h$ both measured from the critical point as~\cite{YamajiLifshitz} 
   \begin{equation}
    F = -\Delta h\Delta m + v(\Delta m)^2 + c_2(\Delta m)^{(d+2)/2} +\cdots .
    \label{hyperscaling8}
  \end{equation}
Equation (\ref{hyperscaling8}) is obtained from $F_0\propto (\Delta m)^{(d+z)/z}$ instead of Eq.(\ref{hyperscaling2}), because 
of $\Delta m\propto E_F$ and $E_F\propto X^{z/d}$. If a neck of the Fermi surface changes its topology~\cite{YamajiLifshitz}, it is expanded in two dimensions as 
   \begin{equation}
    F = -\Delta h\Delta m + v(\Delta m)^2 + c_{2l}\frac{(\Delta m)^{2}}{\ln \frac{1}{|\Delta m|}} + c_4(\Delta m)^{3} +\cdots ,
    \label{hyperscaling9}
  \end{equation}
and in three dimensions as inferred from Eq.(\ref{hyperscaling8}) as
   \begin{equation}
    F = -\Delta h\Delta m + v(\Delta m)^2 + c_2(\Delta m)^{5/2} + c_3(\Delta m)^{3} +\cdots ,
    \label{hyperscaling10}
  \end{equation}
for the disconnected side of the neck-collapsing transition and 
   \begin{equation}
    F = -\Delta h(\Delta m) + v(\Delta m)^2 + c_3(\Delta m)^{3} +\cdots ,
    \label{hyperscaling11}
  \end{equation}
for the connected side.

%The unconventional feature in two dimensional systems is summarized as follows: At the MQCP for the canonical ensemble, the susceptibility of the ferromagnetic order parameter $\chi$ diverges as $\ln 1/|\delta\Delta|$ when the ``neck'' of the Fermi surface collapses between the disconnected and connected surfaces at the van Hove singularity. More remarkably, it diverges as $|\delta\Delta|^{-1}$ when the electron/hole pocket of the Fermi surface vanishes. Here $\delta\Delta$ is the amplitude of the mean field measured from the Lifshitz critical point. 
%On the other hand, for the grand canonical ensemble, the first-order transitions appear as the electronic phase separation and the endpoint of the phase separation is the MQCP. Especially, when the pockets of the Fermi surface vanish, the uniform charge compressibility $\kappa$ diverges as $|\delta n|^{-1}$, instead of $\chi$, where $\delta n$ is the electron density measured from the critical point. Accordingly, Lifshitz transition induces large charge and spin fluctuations represented by enhanced $\chi$ and/or $\kappa$. 

In fact, this mechanism in three dimensional systems has been proposed to be relevant in unconventional criticality of metamagnetic quantum critical end point for ZrZn$_2$~\cite{YamajiZrZn2}.
Itinerant ferromagnets such as ZrZn$_2$~\cite{Kimura,Uhlarz}, UGe$_2$~\cite{Huxley} and nearly ferromagnetic metals such as Sr$_3$Ru$_2$O$_7$~\cite{Perry} show metamagnetic transitions. The magnetizations show jumps at magnetic fields separating the low-field phase from the high-field phase with a higher magnetic moment.  The first-order transition terminates at a finite-temperature critical point.  The universality around the critical point is again regarded as the Ising type, which is equivalent to the gas-liquid critical points. The critical temperature can, however, be controlled to zero, for example, by pressure, which offers a QCP.  A possible connection of these QCPs to non-Fermi-liquid behavior as well as superconductivity found in UGe$_2$ stimulated extensive studies~\cite{Saxena}.  
%Especially in weak itinerant-electron  ferromagnet ${\rm ZrZn}_{2}$, the metamagnetic transition is analyzed in terms of the tunability to the quantum critical end point~\cite{Kimura,Uhlarz}. A pair of nearly quantum metamagnetic transitions appears in a bilayered ruthenate ${\rm Sr_{3}Ru_{2}O_{7}}$ under a high magnetic field around 8 T~\cite{Perry,Grigera,Green}, which has attracted much attention in relation to electronic phase separations and hidden orders~\cite{phaseseparation}.

%A simple scope of the QCP is offered by suppressions of critical temperatures of gas-liquid (or equivalently Ising) transitions, $T_{c}$, by quantum fluctuations.  
\begin{figure}[htb]
\begin{center}
\includegraphics[width=8cm]{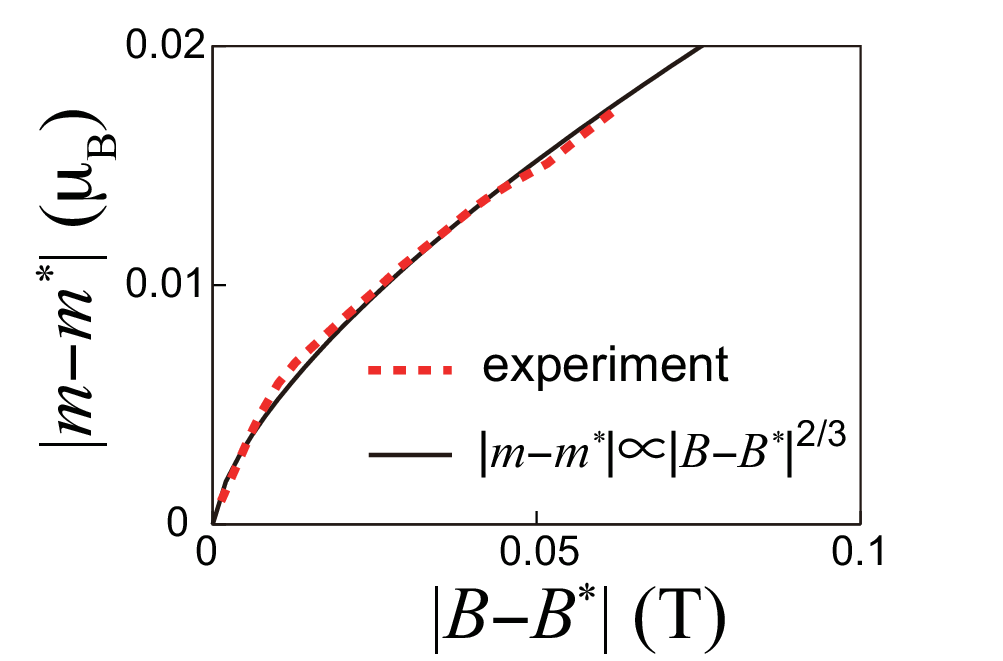}   
\caption{Magnetic field dependence of magnetization, for both measured from the critical point. Theoretical prediction~\cite{YamajiZrZn2} of the MQCP plotted as solid (black) curve reproduces the experimental results for ZrZn$_2$ given by the dashed (red) curve~\cite{Uhlarz}. This is evidence for the MQCP described by $\delta =3/2$}
\end{center}
\label{Fig4}
\end{figure}%
These metamagnetic transitions have first been analyzed by the conventional framework of the quantum criticality of symmetry breaking~\cite{Millis02}.
However, it has been proposed that the topological change in the Fermi surface as the neck-collapsing type is responsible for the metamagnetic behavior for ZrZn$_2$ on the basis of the analyses by the band structure calculation~\cite{YamajiZrZn2}.
Then the free energy has the form of Eqs. (\ref{hyperscaling10}) and (\ref{hyperscaling11}). 
In this case, the critical exponent $\delta$ defined by
$\Delta m \propto |\Delta h|^{1/\delta}$ is given by
$\delta=3/2$ for the side of the disconnected neck and $\delta=2$ for the side of the connected neck. This is in sharp contrast with
the Ising universality value $\delta \sim 4.8$. It is largely different even from the Ising mean-field value $\delta=3$.
This exponent predicts a convex curve for the inverse of uniform magnetic susceptibility $\chi_0^{-1}$ as a function of magnetization as $\chi_0^{-1}\propto |\Delta m|^{1/2}$ in the disconnected side.  This remarkable feature is consistent with the experimental indications by  Uhlarz {\it et al}.~\cite{Uhlarz} as illustrated in Fig.~\ref{Fig4}.

\section{Discussions}
We have discussed mechanisms of several unconventional quantum criticalities
associated with the proximity to first-order transitions.
The first case is the quantum tricriticality in metals, where the conventional
theory of quantum criticality for symmetry-breaking transitions is
substantially modified by the coupling of three low-energy modes, namely,
uniform excitations, the order parameter, and quasiparticle excitations. 
The second case is topological transitions of Fermi surface coupled to electron correlations,
including metal-insulator transitions and Lifshitz transitions. 

In all the cases, the proximity
is a source of the unconventional non-Fermi liquids.  The quantum tricriticality generates a crossover region
of non-Fermi liquids at nonzero temperatures while Fermi liquids are recovered at sufficiently low temperatures,
except for the exact QTCP.  
On the other hand, if the symmetry breaking is suppressed, the continuous topological transitions
accompany an extended area of non-Fermi-liquid phase caused by the zeros of Green function.
Although the MQCP generates a cone-shape structure of the critical region similarly to the 
conventional quantum criticality as we saw in Fig.~\ref{Fig1}(b), the unconventional metals 
have wider extension as a phase because of the electron differentiation caused by the penetrating zeros. 
In this respect, physics of the metal-insulator and Lifshitz transitions 
waits for further studies not only on the criticality but also on the extended non-Fermi-liquid phase.
First-order transitions may also introduce an extended spatially inhomogeneous region in the parameter space
including phase separations. 
Unexpectedly wide regions of non-Fermi liquids recently pointed out in various experiments
may have a connection to this topological aspect combined with the proximity to the first-order transition~\cite{Anderson}.
Another intriguing issue is to elucidate the universality of the possible MQCP for other topological transitions such as transitions of quantum Hall states and topological insulators.

\ack
The authors thank Yukitoshi Motome and Shiro Sakai for useful discussions and collaborations in a part of this work. MI is also grateful to Phil W. Anderson for fruitful discussions on Ref.\cite{Anderson}. This work has been supported by
Grant-in-Aids from MEXT Japan.

\end{document}